\documentclass[%
reprint,
superscriptaddress,
prl,
 amsmath,amssymb,
 aps,
]{revtex4-2}
\usepackage{color}
\usepackage{gensymb}
\usepackage{graphicx}
\usepackage{dcolumn}
\usepackage{bm}
\usepackage{hyperref}
\usepackage{epstopdf}
\usepackage{float}
\usepackage{listings}
\usepackage{times,mathptm,bm}
\usepackage{xcolor}
\usepackage[export]{adjustbox}
\usepackage{mhchem}
\usepackage{siunitx}
\usepackage{booktabs}

\begin{document}

\title{The Paper-Stack Trampoline: Acoustic Restitution in Layered Media}

\author{Manou~Liesker} 
\affiliation{Van der Waals-Zeeman Institute, Institute of Physics, University of Amsterdam, 1098XH Amsterdam, The Netherlands.}
\author{Colton~Kawamura}
\affiliation{Van der Waals-Zeeman Institute, Institute of Physics, University of Amsterdam, 1098XH Amsterdam, The Netherlands.}
\author{Maria~Kieft}
\affiliation{Van der Waals-Zeeman Institute, Institute of Physics, University of Amsterdam, 1098XH Amsterdam, The Netherlands.}
\author{Joshua~A.~Dijksman}
\email{j.a.dijksman@uva.nl}
\affiliation{Van der Waals-Zeeman Institute, Institute of Physics, University of Amsterdam, 1098XH Amsterdam, The Netherlands.}
\author{Antoine~Deblais}
\email{a.deblais@uva.nl}
\affiliation{Van der Waals-Zeeman Institute, Institute of Physics, University of Amsterdam, 1098XH Amsterdam, The Netherlands.}

\date{\today}

\begin{abstract}
A ball bouncing on a thicker cushion should bounce less. We show that this is not always so . Measuring the coefficient of restitution $e$ of a steel ball dropped on a stack of $N$ sheets of standard A4 printer paper, we find that $e$ first drops, then rises to a pronounced maximum, and only then decreases towards a minimum. The maximum occurs where the acoustic round trip through the stack matches the contact time of the impact. For a contact set by the stack itself, one recovers the classical bar-impact condition of a stack-to-ball mass ratio of order one. Bar impact, however, predicts only a weak recovery, insensitive to the surrounding gas. For paper stacks, the recovery is large and no maximum is resolved when the stack is evacuated: the air trapped between the sheets, not the paper alone, makes the stack a trampoline.
\end{abstract}

\maketitle

Drop a ball on a hard floor and it bounces; slide a stack of paper underneath and it bounces less. This is the expected behavior of any cushion: adding dissipative material between the impactor and the ground should degrade the rebound monotonically. Here we report that a stack of ordinary printer paper violates this expectation. As sheets are added, the coefficient of restitution $e$ of a steel ball first decreases, as expected, but then \emph{increases} again over a broad range of stack heights, reaching a maximum before settling onto a plateau. There is a number of sheets at which a stack of paper is the best trampoline.

The coefficient of restitution, defined as the ratio of velocities before and after impact, is an old  and crude summary of an impact~\cite{Stronge2000,Goldsmith,Johnson}. Predicting it from material properties has nevertheless resisted a century of effort, because $e$ aggregates plastic yielding, viscoelastic relaxation, contact mechanics, porosity, and the radiation of elastic waves into a single number~\cite{Hertz,Raman1920,Hunter1957,Reed1985,Gorin2025}. The radiative dissipation mechanism has been thoroughly studied. Zener showed in 1941 that a sphere striking a large plate loses energy to flexural waves carrying momentum away from the contact~\cite{Zener1941}. Radiation is, however, only irreversible if the waves never come back. Its one-dimensional archetype is the classical Saint-Venant problem of a mass striking a finite elastic bar~\cite{TimoshenkoGoodier,Stronge2000}: the stress wave reflects from the far end and, depending on the bar-to-impactor mass ratio, returns momentum to the impactor or remains trapped in the bar, the same timing competition behind the ``trampoline effect'' of sports implements~\cite{Cross1999} and the bounce of bead columns~\cite{Falcon1998}. Experiments on spheres bouncing from plates of decreasing thickness have accordingly identified a critical thickness below which wave losses collapse~\cite{Sondergaard1990,Tillett1954,Patil2017}. What has been missing is a system in which the competition between wave recovery and loss can be tuned continuously in a single experiment.

A stack of paper is such a system, for a reason that is not obvious. As we shall see, waves in the stack propagate through a chain of rough, weakly loaded sheet--sheet contacts separated by thin layers of trapped air. Both ingredients are extremely compliant, and the resulting medium belongs to the family of slow acoustic materials such as granular packings~\cite{LiuNagel1992,Jia1999}, bubbly liquids~\cite{Wood1930,Bretagne2011} and soft porous solids~\cite{Brunet2013,Ba2017}, in which effective sound speeds far below those of either constituent are routine. Unlike a simple bar, such a medium can also store energy in fluid trapped in the microstructure, and dissipate it through squeeze flow and friction. Paper stacks have served as model systems for the mechanics of complex assemblies, most famously in the self-amplification of friction between interleaved books~\cite{Alarcon2016,DalnokiVeress2016} and its relatives in entangled granular matter and elongated fibers~\cite{Dumont2018,Seguin2022}; their impact response, by contrast, has received little attention.

\begin{figure*}[t]
\includegraphics[width=1.9\columnwidth]{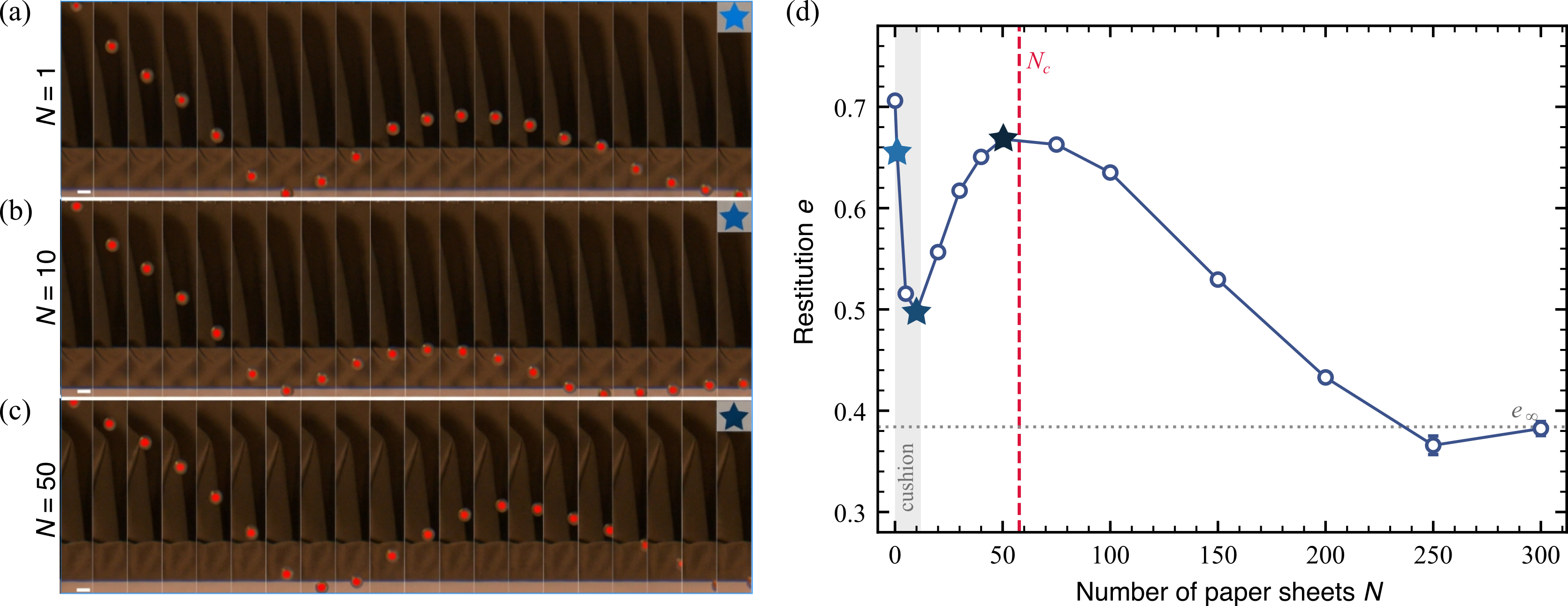}
\caption{\label{fig:cor80}
The trampoline effect of a paper stack. Representative sequences of pictures of a steel ball, initially dropped from the same height, bouncing on a standard printer paper stack of density $80\,\mathrm{g\,m^{-2}}$ for three different numbers of sheets: (a)~$N=1$, (b)~$N=10$, and (c)~$N=50$, chosen respectively during the initial decrease, at the minimum, and near the maximum. The red dot indicates the tracked centroid of the ball; frames are separated by $25\,\mathrm{ms}$; scale bar $15\,\mathrm{mm}$.
(d)~Restitution coefficient of the corresponding paper stack, as a function of the number of sheets $N$. Instead of decreasing monotonically, $e$ falls over the first few sheets (shaded: the ordinary cushioning regime), recovers over a broad maximum centered at $N\simeq58$, and saturates at $e_\infty\simeq0.39$. Stars mark the three stacks shown in (a)--(c). Dashed line: the matching condition, Eq.~(\ref{eq:Nc}). Error bars are standard errors over at least three drops, and are smaller than the symbols in most cases.}
\end{figure*}

In this Letter we show that the maximum in $e(N)$ occurs when the acoustic round-trip time across the stack matches the contact time of the impact, consistent with the Saint-Venant work, but that the magnitude of $e$ is set by the air trapped between the sheets, and can be removed entirely. The main condition is that the ball is still touching the stack when the wave it launched returns, and the returning pulse adds to its upward momentum at separation. Maximal restitution then occurs at a critical number of sheets
\begin{equation}
N_c=\frac{c\,\tau}{2d},
\label{eq:Nc}
\end{equation}
where $c$ is the compression-wave speed in the stack, $\tau$ the duration of the compression phase of the impact, and $d$ the effective thickness of one paper--air cell. Expressing $N$ in units of $N_c$ yields the dimensionless control parameter
\begin{equation}
\Lambda\equiv\frac{2Nd}{c\,\tau}=\frac{N}{N_c},
\label{eq:Lambda}
\end{equation}
the acoustic round-trip time measured in units of the contact time. For a steel ball on a compliant stack, $\tau$ is itself set by the stack, and Eq.~(\ref{eq:Nc}) becomes a condition on the stack-to-ball mass ratio, as we show below.

In our experiment, a steel sphere of diameter $2R=15\,\mathrm{mm}$ and mass $m=13.71\,\mathrm{g}$ is held at a fixed height $H=20\,\mathrm{cm}$ by a $3$D-printed soft vacuum gripper, and released without rotation by venting the gripper (see Supplemental Material~\cite{Sup}), giving an impact velocity $v_i=\sqrt{2gH}=1.98\,\mathrm{m\,s^{-1}}$. The sphere falls onto a stack of $N$ identical sheets resting on a rigid substrate; experiments are in ambient air unless otherwise noted. The trajectory is recorded at $200\,\mathrm{fps}$; the light source, positioned behind the camera, produces a bright specular spot on the sphere that is tracked frame by frame, and the parabolic free-flight arcs before and after impact are fitted separately to extract the drop and rebound apex heights $H$ and $h$. Since dissipation is confined to the impact and air drag is negligible over the flight, $e=v_f/v_i=\sqrt{h/H}$. Each point is the mean of at least three drops and error bars are standard errors of the mean. Five paper grades were used, with areal density $\sigma=40$, $80$, $120$, $160$ and $250\,\mathrm{g\,m^{-2}}$, with $N$ from $1$ to $500$. All drops were made at the center of the A4 sheets; the restitution depends measurably on the distance to the nearest edge for thick stacks~\cite{Sup}, and all data below were taken at this position.

Figure~\ref{fig:cor80} shows the main finding. Three regimes are visible in $e(N)$ for the $80\,\mathrm{g\,m^{-2}}$ paper. From $0.71$ on the bare substrate, $e$ drops steeply to $0.50$ at $N=10$, the expected signature of a growing dissipative cushion. Beyond $N\approx10$ the trend reverses and $e$ climbs back over several tens of sheets to a broad maximum, recovering four fifths of the loss [Fig.~\ref{fig:cor80}(d)]. Fitting the centroid of the region within $1.5\%$ of the maximum gives $N_{\max}=58$ for this grade. For larger $N$ the restitution decays again and saturates at $e_\infty\simeq0.39$.

\begin{figure}[b]
\includegraphics[width=0.8\columnwidth]{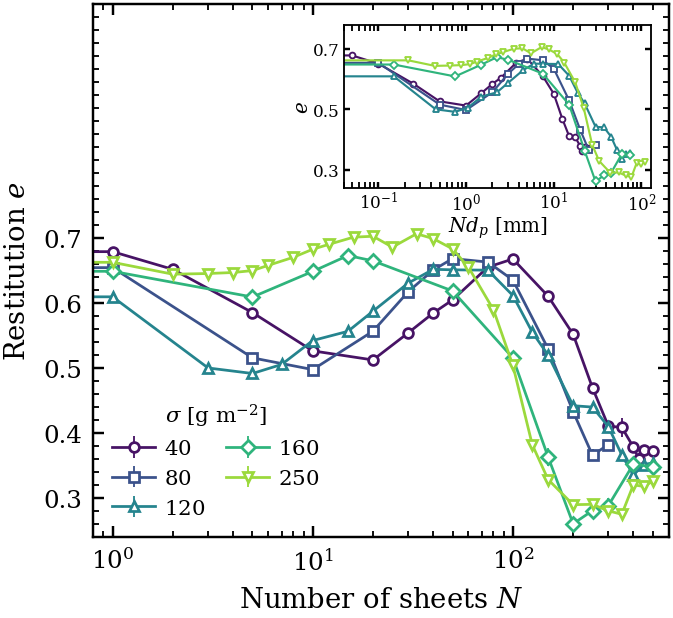}
\caption{\label{fig:allgrades}
Effect of paper density on the restitution coefficient. $e(N)$ for five paper grades, $\sigma=40$ to $250\,\mathrm{g\,m^{-2}}$. All grades are
non-monotonic; the position and height of the maximum depend on the grade. Inset: the same
data against nominal paper thickness $Nd_p$, showing partial collapse of the three
lightest grades.}
\end{figure}

The effect is far outside the scatter and reproduces across independently built stacks and across paper grades (Fig.~\ref{fig:allgrades}). All grades show a maximum at intermediate $N$, with $N_{\max}=100$, $58$, $55$, $18$ and $27$ for $\sigma=40$ to $250\,\mathrm{g\,m^{-2}}$, and $e_\infty$ between $0.32$ and $0.39$; the initial minimum is weaker for the heaviest grade. The maximum moves by a factor of five with grade, while the corresponding stack \emph{height} $N_{\max}d$ varies much less, from $6.5$ to $9.1\,\mathrm{mm}$ for the three lightest grades. Re-plotting against the nominal paper thickness $Nd_p$ (inset) brings the $40$, $80$ and $120\,\mathrm{g\,m^{-2}}$ curves close together but leaves the two heaviest grades apart.

In fact, the stack is not simply a pile of $N$ sheets: stacking $N$ sheets does not produce a height $Nd_p$. Measuring the free-stack height with a CMOS laser distance sensor (FSD22-100P-UI~\cite{Sup}) gives a height growing linearly with $N$, confirming that the stack is statistically uniform, but with a slope $d_{80}=114\,\mathrm{\mu m}$ exceeding the caliper thickness $d_p=100\,\mathrm{\mu m}$ of a single sheet; the $120\,\mathrm{g\,m^{-2}}$ paper gives $d_{120}=165\,\mathrm{\mu m}$ against $d_p=150\,\mathrm{\mu m}$. The excess, $d_a=d-d_p\approx15\,\mathrm{\mu m}$, is air trapped between neighboring sheets by surface roughness, porosity and fiber-scale undulations. It is expelled by any contact probe, which is why the measurement must be made without loading the stack; it adds $9$--$13\%$ to the stack height for both grades measured~\cite{Sup}. The repeating unit is therefore not a sheet but a \emph{paper--air cell} of thickness $d=d_p+d_a$, and it is this composite cell, far more compliant than the paper alone, that carries the compression wave.

\emph{Mechanism.} The lateral dimensions of the sheets exceed the stack height by two orders of magnitude, so on the timescale of the impact the stack is a one-dimensional layered medium of $N$ paper--air cells of thickness $d$, bounded above by the free surface and below by a rigid wall. An impact event compresses the topmost cells and launches a longitudinal pulse downwards. The pulse reflects from the substrate (in phase, the substrate being acoustically much stiffer than the stack) and returns to the free surface after
\begin{equation}
t_r=\frac{2Nd}{c}.
\label{eq:tr}
\end{equation}
Meanwhile the contact has a finite lifetime: the stack is compressed for a time $\tau$, reaches maximum compression, then decompresses as the ball is expelled. We can define $\tau$ as half the total contact time.

The competition between $t_r$ and $\tau$ organizes the whole curve. For small $N$ the first sheets add a soft, lossy contact layer (indentation and friction of the top sheets) without enough thickness to return energy, so $e$ drops below its bare-substrate value. For $t_r\gg\tau$ ($\Lambda\gg1$) the ball has left before any reflection arrives; the radiated energy is lost to the stack, and $e$ becomes independent of $N$, with a plateau $e_\infty$. Between the two lies the matched case $t_r\simeq\tau$, an acoustic resonance of the impact, where the reflected pulse returns as the stack reaches maximum compression and pushes the ball throughout the unloading phase. Its momentum is delivered to the ball rather than dissipated in the stack, and $e$ is maximal: at resonance the ball recovers nearly as much rebound as from the bare substrate itself [Fig.~\ref{fig:cor80}(d)]. Setting $t_r=\tau$ in Eq.~\eqref{eq:tr} yields Eq.~\eqref{eq:Nc}.

\begin{figure}[t]
\includegraphics[width=0.8\columnwidth]{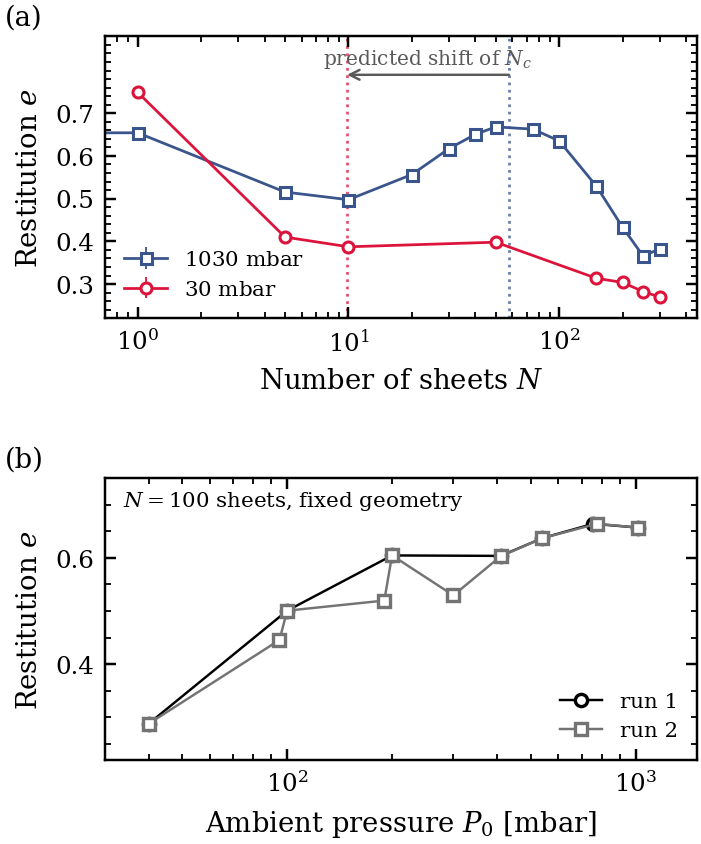}
\caption{\label{fig:pressure}%
Effect of the trapped air on the restitution coefficient. All data for $80\,\mathrm{g\,m^{-2}}$ paper.
(a)~$e(N)$ at atmospheric pressure and at $30\,\mathrm{mbar}$. Dotted lines: $N_c$ from Eq.~(\ref{eq:Nc}) assuming a fixed $\tau$ and $c\propto\sqrt{P_0}$, shown for reference; in the bar picture the maximum would instead stay near $N\simeq50$. Neither is observed at $30\,\mathrm{mbar}$.
(b)~Restitution at fixed geometry, $N=100$ sheets, as a function of ambient pressure, from two independent runs.}
\end{figure}

Which mass of the paper stack is relevant? For a steel ball on a stack of effective modulus $\rho c^2\approx50\,\mathrm{kPa}$, the ball is effectively rigid and the contact time is set by the stack. Modeling the loaded column as a bar of total mass $M_c=N\sigma A$ and stiffness $k=\rho c^2A/Nd$, with $A$ the loaded area, $\tau\simeq(\pi/2)\sqrt{m/k}$ and Eq.~(\ref{eq:Lambda}) reduces to $\Lambda\simeq(4/\pi)\sqrt{M_c/m}$, independent of $c$: the matching condition is the Saint-Venant bar-impact condition $M_c/m\simeq\pi^2/16\approx0.6$~\cite{TimoshenkoGoodier,Stronge2000}. Applied to the measured $N_{\max}$, it gives a loaded radius between $2$ and $3\,\mathrm{cm}$ for all five grades, a few ball diameters, as expected from the lateral load spreading of the sheets.

For consistency, the timing condition can also be checked directly against independently measured $d$, $c$ and $\tau$. The wave speed $c$ is obtained by placing the stack directly on a force sensor and impacting it (FlexiForce, details in~\cite{Sup}). Beyond the primary impact peak the traces show a secondary feature, delayed by a few milliseconds, which we identify as the echo of the compression pulse reflecting from the rigid base of the sensor; the bare sensor shows no comparable feature~\cite{Sup}. This gives $c_{80}\approx8.3\,\mathrm{m\,s^{-1}}$ and $c_{120}\approx10\,\mathrm{m\,s^{-1}}$. The contact time is measured optically, from an impact recorded at $100\,000$~fps (Phantom TMX 7510, Vision Research~\cite{Sup}), giving a total contact time of $3.0\,\mathrm{ms}$ and hence $\tau\simeq1.5\,\mathrm{ms}$. With these values Eq.~\eqref{eq:Nc} predicts $N_c^{80}=55$ and $N_c^{120}=50$, against observed maxima at $N_{\max}=58$ and $55$; equivalently, $t_r$ evaluated at the observed maximum is $1.59$ and $1.82\,\mathrm{ms}$, against $\tau\simeq1.5\,\mathrm{ms}$. Given that $\tau$ comes from a single impact and the echo is a broad feature, this agreement holds within the experimental uncertainty. In the bar picture it is expected: $t_r\simeq\tau$ and $M_c/m\simeq0.6$ are the same statement.

A numerical one-dimensional ball--chain model, in which each paper--air cell is a linear spring--dashpot and a heavy impactor of mass ratio $\hat{m}$ strikes the chain, serves as a discretized bar~\cite{Sup}. It reproduces the drop, maximum and decay of Fig.~\ref{fig:cor80}(d), and its maximum tracks the bar prediction $(N-1)\simeq0.62\,\hat{m}$, confirming that the optimum sits at $t_r\simeq\tau$ rather than $t_r\simeq2\tau$~\cite{Sup}. Its recovery, however, amounts to only a few percent in $e$, against four fifths of the loss in the experiment. Bar impact theory thus accounts for where the maximum lies, but neither for its size nor, more importantly, for why the recovery should depend so strongly on the medium between the layers. This distinction motivates a direct test of the role of the trapped gas.

\emph{The role of the trapped air.} Ambient pressure discriminates a stack of  paper from a bar. Lowering $P_0$ can e.g. soften the pockets of air between the paper layers; in the bar picture this lowers $c$ but leaves $\Lambda$, and hence the maximum and its small amplitude, unchanged. To test this picture, we place the same setup inside a transparent acrylic box (Sanatron) connected to a vacuum pump, lowering the pressure to $30\,\mathrm{mbar}$. Because of space constraints the rigid substrate is here the acrylic wall rather than the wooden substrate used at ambient pressure, so the ball--substrate restitution at $N=0$ differs slightly between the two conditions [Fig.~\ref{fig:pressure}(a)]. For $N=100$ sheets, $e$ falls from $0.66$ at atmospheric pressure to $0.29$ at $40\,\mathrm{mbar}$ [Fig.~\ref{fig:pressure}(b)], reproducibly over two independent runs, well below the $1\,\mathrm{atm}$ plateau. At $30\,\mathrm{mbar}$ the full $N$ scan shows no resolved maximum: $e=0.41$, $0.39$ and $0.40$ at $N=5$, $10$ and $50$, where bar theory places the maximum [Fig.~\ref{fig:pressure}(a)]. Without air, the stack behaves as a lossy bar with no appreciable recovery; with air, the recovery is large.

The air cannot act as a simple sealed spring. A fully trapped, isothermal gas layer has stiffness per unit area $P_0/d_a$ and would raise the cell modulus to $P_0d/d_a\approx0.8\,\mathrm{MPa}$, i.e. $c\approx33\,\mathrm{m\,s^{-1}}$, four times the measured value. Over the contact footprint $a\simeq R$, we can define a ``squeeze'' number $S=12\eta\omega a^{2}/P_0d_a^{2}\approx1$, so the gas partially vents during the impact: it acts as a squeeze-film element that both stores and dissipates energy on the timescale of the contact, and may also spread the load laterally, a channel absent from bar theory. Its viscous shear stress ($\simeq5\,\mathrm{Pa}$) is negligible compared with $P_0$.

That the air must vent laterally is confirmed directly: the restitution depends on where on the sheet the ball lands~\cite{Sup}. For thin stacks ($N=5$, $10$) $e$ is independent of the distance from the edge, whereas for $N=100$ it rises from $0.49$ to $0.63$ between $2$ and $15\,\mathrm{cm}$, and for $N=300$ from $0.29$ to $0.46$ beyond $10\,\mathrm{cm}$. Impacts near a free edge, where trapped gas escapes most easily, are the least elastic.

\begin{figure}[t]
\includegraphics[width=0.8\columnwidth]{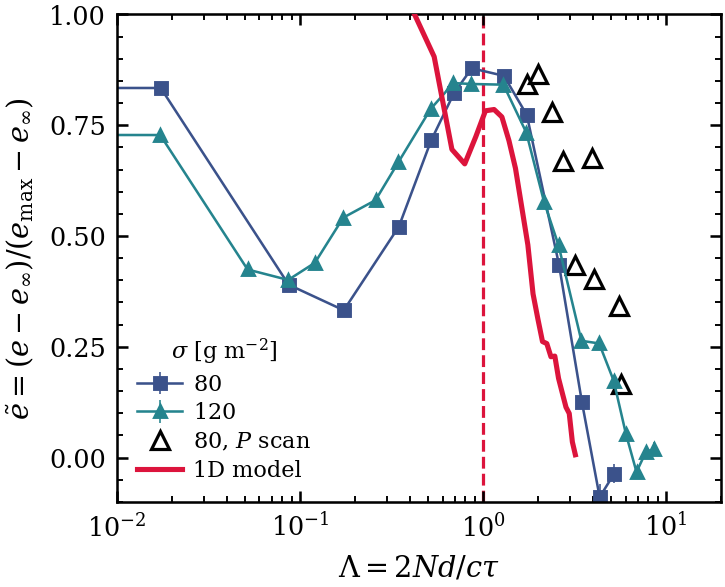}
\caption{\label{fig:collapse}%
Normalized restitution $\tilde{e}=(e-e_\infty)/(e_{\max}-e_\infty)$ versus the dimensionless round-trip time $\Lambda=2Nd/c\tau$, for the two grades with measured $d$ and $c$, together with the fixed-$N$ pressure scan mapped onto the same axis assuming $c\propto\sqrt{P_0}$. The layer scans collapse onto a single curve with a maximum at $\Lambda\simeq1$. The red line is the one-dimensional ball--chain model~\cite{Sup}, with its $\Lambda$ axis rescaled by a constant factor since its contact time differs from the measured one.}
\end{figure}

Figure~\ref{fig:collapse} shows $\tilde{e}=(e-e_\infty)/(e_{\max}-e_\infty)$ against $\Lambda$ for the two grades with measured $d$ and $c$: they collapse, with a maximum near $\Lambda=1$. The pressure scan, mapped onto $\Lambda$ assuming a fixed $\tau$ and $c\propto\sqrt{P_0}$, does not perfectly collapse and lies systematically above, perhaps consistent with pressure acting on storage and dissipation rather than on timing alone.

\emph{Discussion.} The mechanism requires that an acoustic round trip in the target be comparable with the contact time, and it is notable that this is achievable with a table-top object at all. For a steel plate, $c\approx5\times10^{3}\,\mathrm{m\,s^{-1}}$ and $\tau\approx10^{-4}\,\mathrm{s}$ put the matched condition at a quarter of a meter of steel, which is why restitution on solid targets is described either quasistatically or as pure radiative loss. For a homogeneous elastic sphere, the Hertzian contact time and the internal round trip $2R/c$ give $\Lambda\approx0.8(v/c)^{1/5}$, independent of size: about $0.2$ for steel and $0.4$ for rubber at $2\,\mathrm{m\,s^{-1}}$, and $\Lambda\sim1$ would require impacts faster than sound. Homogeneous bodies therefore always sit in the quasistatic regime, which is why Hertz's theory works; layered, low-impedance assemblies are the route to $\Lambda\sim1$ at everyday speeds.

The paper stack goes beyond a bar in an additional, more interesting way: the position of its maximum follows the classical mass-ratio condition, but its size is set by the air trapped between the sheets, which stores and dissipates energy on the impact timescale. The restitution of a layered material is set not only by how much energy it absorbs, but also by \emph{when} it gives back what it stored, and through which channel. Since these channels are governed by contact compliance and trapped fluid rather than by the bulk moduli of the constituents, they can be engineered, by mass, by surface roughness, by fluid viscosity, offering a route to layered materials whose impact response is tuned through structure rather than composition, in sports gear, packaging, and protective layers. Measuring the contact time and first-arrival time as functions of $N$ and $P_0$ will identify which of these channels dominates.

\begin{acknowledgments}
We thank the Technology Center of the University of Amsterdam for technical support. The students Laura Nijhuis, Mano Tellegen, Noor Witte, and Saladin Shah, from the 1st year BSc course ``Project'' in the UvA/VU Physics \& Astronomy program, are also acknowledged for contributing to the experiments and to the thinking behind the problem.
\end{acknowledgments}

%

\clearpage

\onecolumngrid 

\setcounter{secnumdepth}{3}
\setcounter{section}{0}
\setcounter{subsection}{0}
\setcounter{equation}{0}
\setcounter{figure}{0}
\setcounter{table}{0}

\renewcommand{\thesection}{S\arabic{section}}
\renewcommand{\theequation}{S\arabic{equation}}
\renewcommand{\thefigure}{S\arabic{figure}}
\renewcommand{\thetable}{S\arabic{table}}

\begin{center}
\textbf{\large \vspace*{1.5mm} Supplementary Material for\\ ``The Paper-Stack Trampoline: Acoustic Restitution in Layered Media''} \\
\end{center}
\vspace*{5mm}

\section{Ball release and imaging}
\label{sec:setup}
\begin{figure}[h!]
\includegraphics[width=\columnwidth]{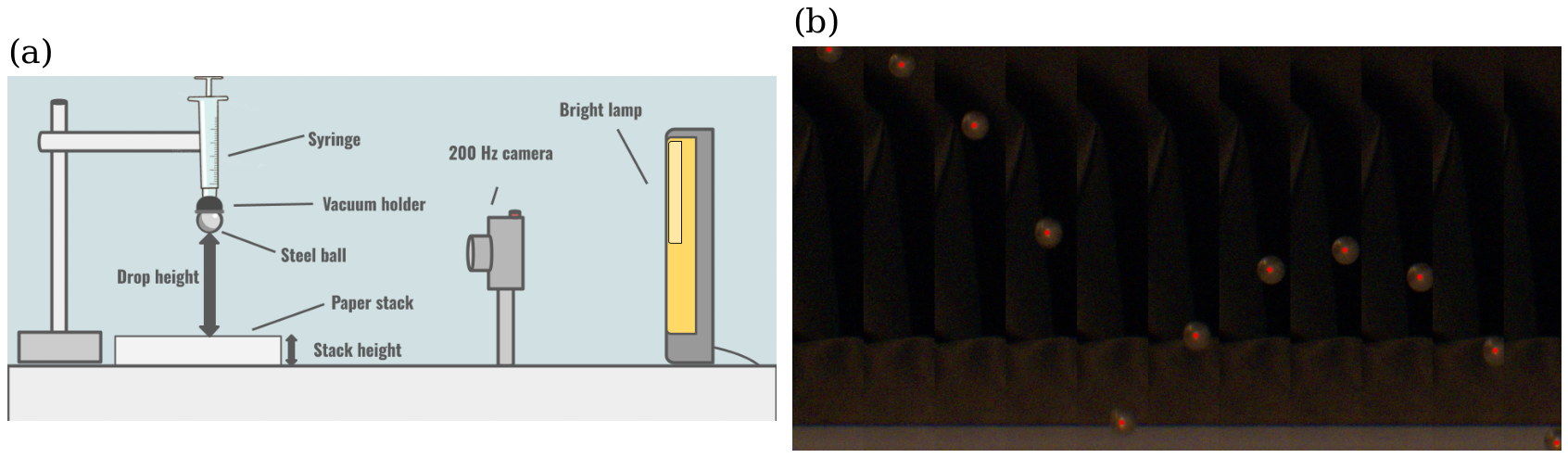}
\caption{\label{fig:setup}%
(a) Schematic of the release and imaging apparatus: the ball is held by a vacuum holder at a fixed drop height above the paper stack, and the fall and rebound are recorded with a camera at $200\,\mathrm{fps}$ under bright, off-axis illumination.
(b) Example sequence of tracked frames: the bright specular reflection on the sphere (tracked point marked in red) is used to determine the trajectory before and after impact, from which the drop and rebound heights $H$ and $h$, and hence $e=\sqrt{h/H}$, are obtained.}
\end{figure}

The steel sphere ($2R=15\,\mathrm{mm}$, $m=13.71\,\mathrm{g}$) is held by suction against a soft, $3$D-printed vacuum gripper mounted on a fixed vertical rail at height $H=20\,\mathrm{cm}$ above the stack [Fig.~\ref{fig:setup}(a)]. Venting the gripper releases the ball without imparting spin or lateral velocity; no measurable rotation is visible in the recorded trajectories. The stack rests on a rigid substrate whose position is fixed for a given series of drops. A bright lamp is positioned to create a specular reflection spot on the sphere, and the fall and rebound are recorded with a camera at $200\,\mathrm{fps}$. This reflection spot is tracked frame by frame [Fig.~\ref{fig:setup}(b)]; the upward and downward flight segments are each fitted to a parabola, and the drop and rebound apex heights $H$ and $h$ are taken from the fitted vertices rather than from individual frames, which is more robust to the finite frame spacing. Each reported point in the main text is the mean of at least three independent drops on independently rebuilt stacks, and the quoted error bars are the standard error of the mean.

All drops discussed in the main text were made at the center of the $\mathrm{A4}$ sheet; the restitution depends measurably on the distance to the nearest edge for thick stacks, for the reason given in Sec.~\ref{sec:edge} below.

\section{Determination of the cell thickness $d$}
\label{sec:cellthickness}
\begin{figure}[t]
\includegraphics[width=\columnwidth]{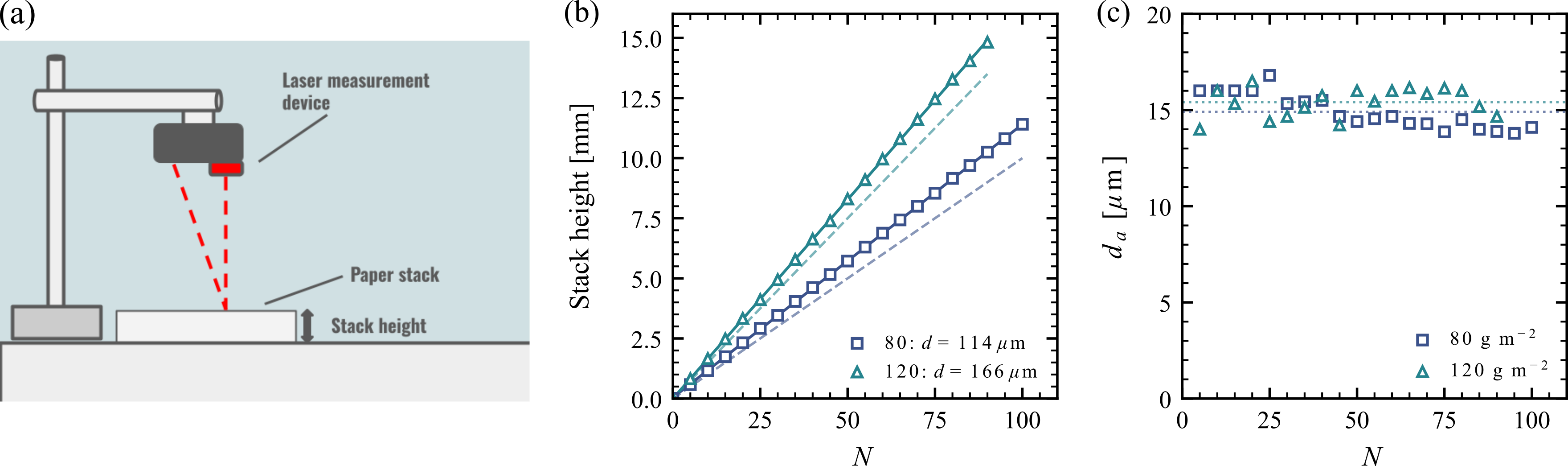}
\caption{\label{fig:cellthickness}
(a) Schematic of the laser-displacement measurement of the free-stack height.
(b) Measured stack height versus $N$ for the $80$ and $120\,\mathrm{g\,m^{-2}}$ papers (solid symbols, linear fits as solid lines) compared with the paper-only estimate $Nd_p$ from the caliper sheet thickness (dashed lines).
(c) Air-gap thickness per layer, $d_a=[\text{height}(N)-Nd_p]/N$, evaluated independently at each measured $N$. Dotted lines: mean value for each grade.}
\end{figure}

The caliper thickness of a single sheet, $d_p$, systematically underestimates the thickness a stack presents to a normal load: surface roughness and fiber-scale undulations trap a thin layer of air between neighboring sheets, and a caliper (or any other contact probe) expels this air before it can be measured. We instead determine the total stack height with a laser displacement sensor [Fig.~\ref{fig:cellthickness}(a)], which does not load the stack mechanically. The height was measured as a function of the number of sheets $N$ for the $80$ and $120\,\mathrm{g\,m^{-2}}$ papers, from $N=0$ to $N=100$ and $N=90$ respectively [Fig.~\ref{fig:cellthickness}(b)]. In both cases the height grows linearly with $N$ over the full range probed, confirming that the effective cell thickness $d=d_p+d_a$ is uniform throughout the stack; the slope of a linear fit gives $d_{80}=114\,\mathrm{\mu m}$ and $d_{120}=165\,\mathrm{\mu m}$, against caliper sheet thicknesses of $d_p=100$ and $150\,\mathrm{\mu m}$. The dashed lines in Fig.~\ref{fig:cellthickness}(b) show the paper-only estimate $Nd_p$ for comparison; the offset between the solid and dashed lines is the trapped air.

The same air-gap thickness can be obtained pointwise, as $d_a(N)=[\text{measured height}(N) - Nd_p]/N$, for every stack independently measured rather than from the slope of a single global fit. This is shown in Fig.~\ref{fig:cellthickness}(c): $d_a$ scatters around $14.9\pm0.9\,\mathrm{\mu m}$ for the $80\,\mathrm{g\,m^{-2}}$ paper and $15.4\pm0.8\,\mathrm{\mu m}$ for the $120\,\mathrm{g\,m^{-2}}$ paper ($1\sigma$ over the measured range of $N$), with no systematic trend with stack size. The two grades agree with each other to within their scatter, so that $d_a\simeq15\,\mathrm{\mu m}$, quoted as a single value in the main text, is representative of both the grade dependence and the $N$ dependence of the trapped-air layer.

\section{Dependence of the restitution on distance from the sheet edge}
\label{sec:edge}
\begin{figure}[t]
\includegraphics[width=0.5\columnwidth]{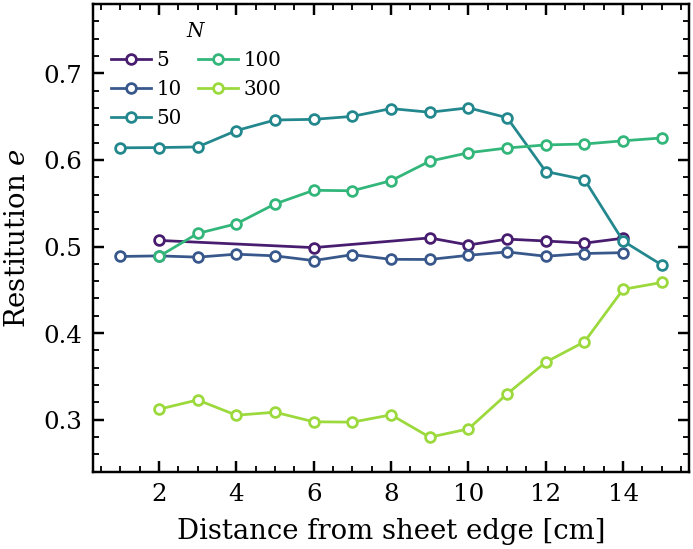}
\caption{\label{fig:edge}
Restitution coefficient as a function of the distance of the impact point from the nearest edge of the stack, for five stack thicknesses $N=5$, $10$, $50$, $100$, and $300$ sheets of $80\,\mathrm{g\,m^{-2}}$ paper.}
\end{figure}

Because the mechanism proposed in the main text requires air trapped between sheets to be compressed on the timescale of the impact, the restitution is expected to depend on how easily that air can escape laterally, and hence on the distance between the impact point and the nearest free edge of the stack. We tested this directly by measuring $e$ as a function of the impact position along the long axis of the sheet, for stacks of $N=5$, $10$, $50$, $100$, and $300$ sheets of $80\,\mathrm{g\,m^{-2}}$ paper (Fig.~\ref{fig:edge}).

For thin stacks ($N=5$ and $N=10$) the restitution is, within scatter, independent of the distance from the edge: the air has little distance to travel regardless of where the ball lands. For $N=100$, $e$ rises from $0.49$ at $2\,\mathrm{cm}$ from the edge to $0.63$ at $15\,\mathrm{cm}$; for $N=300$, $e$ rises from $0.29$ near the edge to $0.46$ beyond $10\,\mathrm{cm}$. In both cases the impact point nearest the free edge, where trapped air can vent most easily, gives the lowest restitution. This confirms that lateral venting of the trapped air is an active process on the timescale of the impact, rather than a static property of the sheet, and it motivates measuring all of the $e(N)$ curves reported in the main text at a single, fixed, central position rather than at the edge.

\section{Contact-time measurement}
\label{sec:contacttime}
\begin{figure}[t]
\includegraphics[width=\columnwidth]{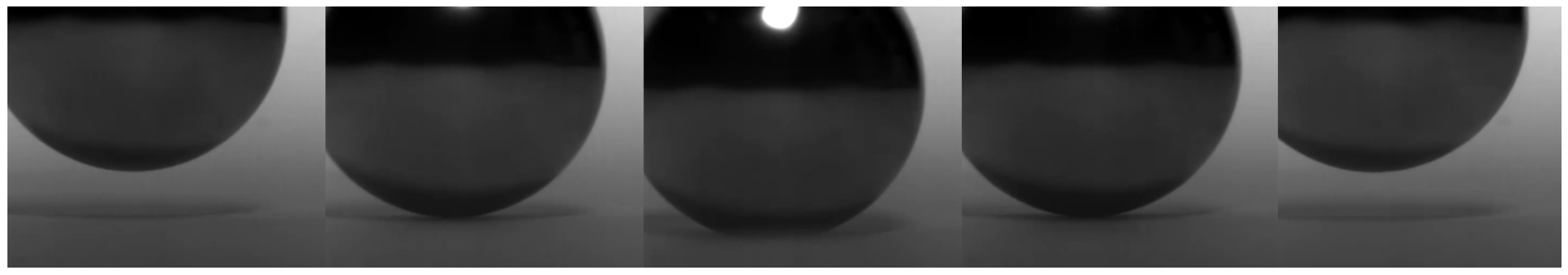}
\caption{\label{fig:100khz}
Consecutive frames from the $100,000\,\mathrm{fps}$ recording used to determine the contact time $\tau$ (Sec.~\ref{sec:contacttime}), for a stack of $N=100$ sheets of $80\,\mathrm{g\,m^{-2}}$ paper. The bright reflection in the central frame marks the point of closest approach. 
Each frame (left to right) shown here are separated by 3 ms. The ball has a diameter 2$R$ = 15 mm.}
\end{figure}

A single impact on a stack of $N=100$ sheets of $80\,\mathrm{g\,m^{-2}}$ paper was recorded at $100\,\mathrm{kHz}$ with a Phantom TMX 7510 high-speed camera (Vision Research). Representative frames spanning the compression of the stack are shown in Fig.~\ref{fig:100khz}; the bright flash visible in the central frame is the specular reflection of the illumination at the point of closest approach, coinciding with maximum compression of the stack. The total duration of contact between the ball and the stack, from first contact to separation, was measured directly from the recorded frames as $2\tau=3.0\,\mathrm{ms}$, giving the compression-phase duration $\tau\simeq1.5\,\mathrm{ms}$ quoted in the main text.

This measurement was performed once, for a single paper grade and a single stack thickness. Section~\ref{sec:wavespeed} shows that the width of the transmitted force pulse measured on the piezoelectric sensor is substantially narrower than this total optical contact time; the two are not the same quantity, since the force sensor registers only the high-amplitude portion of the contact while the ball remains in light contact with the paper for longer, so this is expected rather than a discrepancy in $\tau$ itself. A dependence of $\tau$ on $N$, which would be expected physically once the stack, rather than the ball, sets the contact timescale (see main text), has not yet been measured.

\section{Compression-wave speed}
\label{sec:wavespeed}

\begin{figure}
\includegraphics[width=\columnwidth]{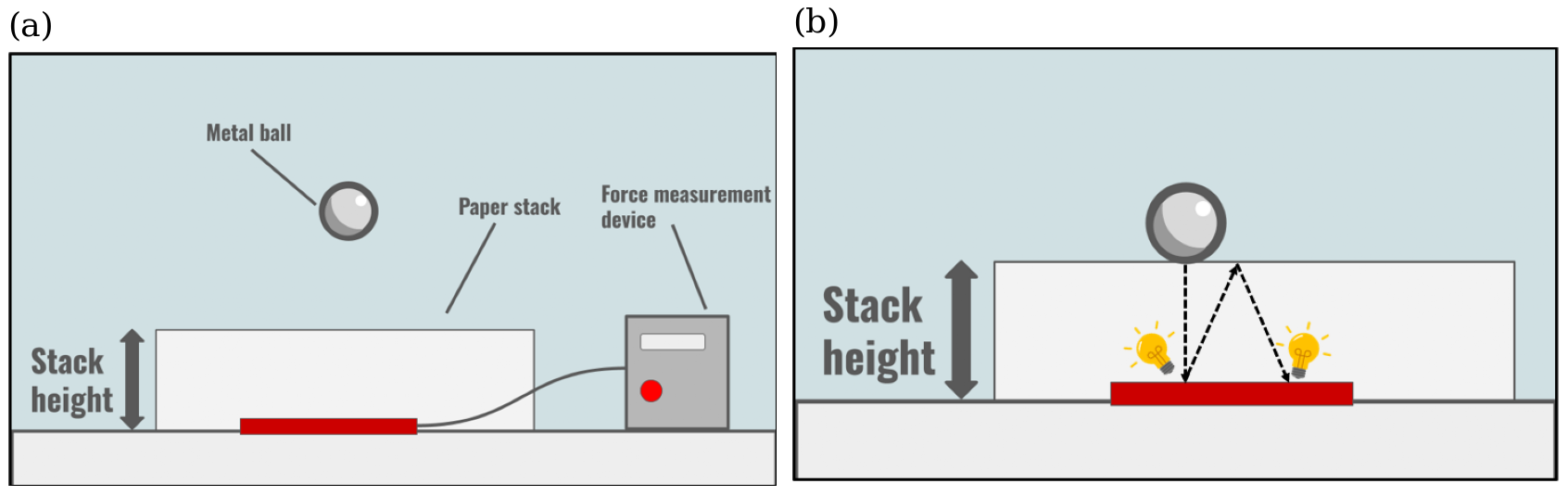}
\caption{\label{fig:wavespeedsetup}
(a) Schematic of the force-sensor measurement: the stack is impacted directly on top of a piezoelectric force sensor.
(b) Candidate mechanism: a compression pulse launched at impact travels down the stack, reflects from the rigid sensor surface (substrate), and returns to the free surface after a round-trip time $t_r=2Nd/c$. The force sensor detects the compressive wave once it has completed this round trip (shown by the bulb lights in the sketch). The corresponding force traces are shown in Sup.~Fig.~S\ref{fig:forcetraces}.}
\end{figure}

\begin{figure}
\includegraphics[width=\columnwidth]{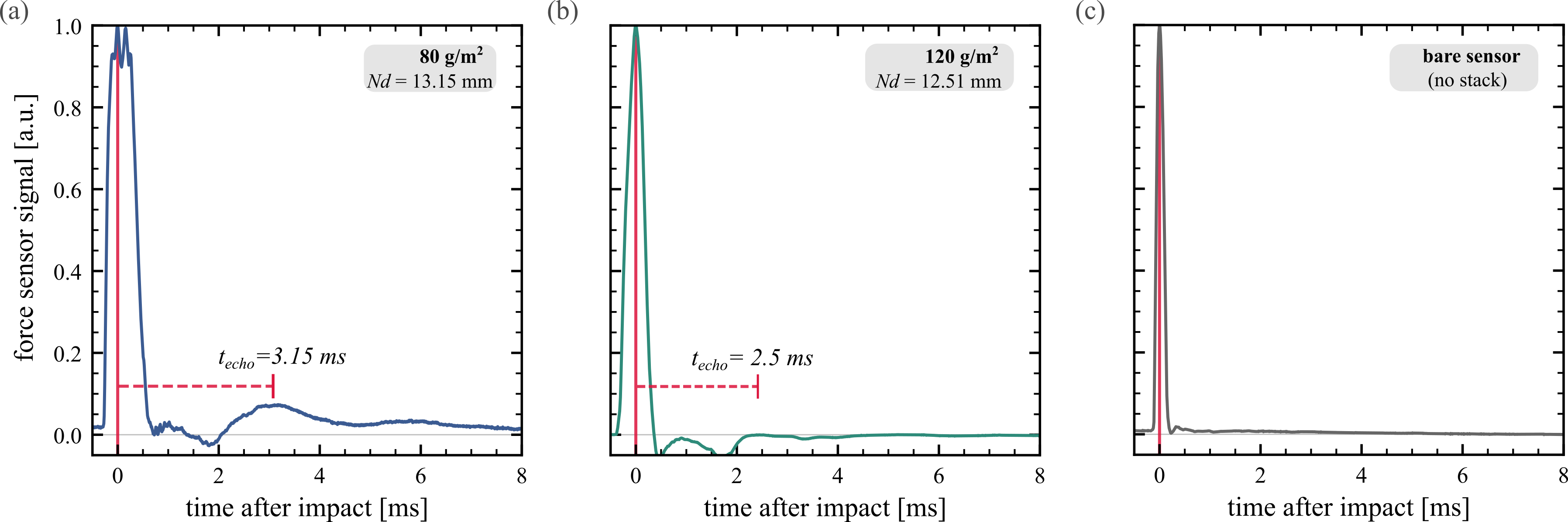}
\caption{\label{fig:forcetraces}
Force-sensor measurements used to estimate the compression-wave speed $c$ (Sec.~\ref{sec:wavespeed}), for $80\,\mathrm{g\,m^{-2}}$ paper ($Nd=13.15\,\mathrm{mm}$), $120\,\mathrm{g\,m^{-2}}$ paper ($Nd=12.51\,\mathrm{mm}$), and a bare-sensor control with no stack present. The solid vertical line marks the primary impact ($t=0$); the dashed vertical line marks the echo, the largest secondary feature in the $0.6$--$8\,\mathrm{ms}$ window. The bare sensor has no notable second peak.}
\end{figure}

The compression-wave speed $c$ was measured by placing a stack directly on a piezoelectric force sensor (FlexiForce, dimensions $5 \times 5\,\mathrm{cm}^2$) and impacting it with the steel sphere [Fig.~\ref{fig:wavespeedsetup}(a)], recording the force--time signal at $400\,\mathrm{kHz}$ (acquisition card from National Instruments, interfaced with a home-made Python script). A compression pulse launched by the impact travels through the stack, reflects off the rigid base at the sensor, and returns to the free surface after a round-trip time $t_r=2Nd/c$ [Fig.~\ref{fig:wavespeedsetup}(b)], which we interpret as a secondary feature in the force signal.

Figure~\ref{fig:forcetraces} shows representative traces for $80$ and $120\,\mathrm{g\,m^{-2}}$ paper, and a bare-sensor control, without a stack on top of it. In each stack trace, a secondary feature at $2$--$10\%$ of the primary peak appears a few milliseconds after impact; we measure $t_{\rm echo}^{80}=3.15\,\mathrm{ms}$ and $t_{\rm echo}^{120}=2.5\,\mathrm{ms}$. Converted to a wave speed via $c=2Nd/t_{\rm echo}$, using the laser-measured stack height $Nd$, this gives $c_{80}=8.3\,\mathrm{m\,s^{-1}}$ and $c_{120}=10.0\,\mathrm{m\,s^{-1}}$, the values used in the main text. The bare sensor shows no comparable feature, confirming that the echo is a property of the stack and not an instrumental artifact.

\section{Ambient-pressure measurements}
The reduced-pressure data shown in the main text were obtained with the stack and ball mounted inside a vacuum chamber, evacuated with a mechanical pump. The $e(N)$ scan at $30\,\mathrm{mbar}$ (main text, Fig.~3) samples $N=1$, $5$, $10$, and $50$ sheets, which is too coarse to resolve the displaced maximum predicted near $N_c\simeq10$ at this pressure; a denser scan over this range would be a direct test of the $c\propto\sqrt{P_0}$ scaling proposed in the main text. The fixed-geometry scan at $N=100$ sheets (main text, Fig.~3) was repeated on two separate days with independently rebuilt stacks, and the two runs agree to within the scatter of repeated drops at fixed $N$.

\section{Idealized one-dimensional ball--chain simulation}
\label{sec:1dmodel}

We reduce the paper--air cell to a single effective spring and dashpot. In the real stack, each cell is a parallel network of rough paper--paper contacts and compliant air gaps, whose stiffness approaches the dry-paper value in vacuum but is strongly softened by trapped air at ambient pressure. At the strains involved here, this microstructure can be collapsed onto one spring constant and one damping coefficient per cell, so that a one-dimensional chain with the appropriate parameters can capture the dominant physics behind $e(N)$.

\subsection{Model definition and non-dimensionalization}

The system is $N$ particles: a single impacting particle (the ``ball'') and a chain of $N-1$ identical particles (the ``chain''). Neighboring particles interact through damped linear springs of rest length $d_c$ and stiffness $k_c$ (chain--chain) or $k_b$ (ball--chain). The bottom particle is fixed at the origin; the ball starts two diameters above the first chain particle with a downward impact velocity. All parameters are expressed in units of the chain diameter $d_c$, mass $m_c$, and spring constant $k_c$ (Table~\ref{tab:nondim}).

\begin{table}[t]
\caption{\label{tab:nondim}
Chain-based non-dimensionalization used in the one-dimensional ball--chain simulation.}
\centering
{%
\setlength{\tabcolsep}{3pt} 
\begin{tabular}{@{}lll@{}}
\toprule
Symbol & Definition & Description \\
\midrule
$d_c$ & chain diameter & length scale \\
$m_c$ & chain mass & mass scale \\
$k_c$ & chain spring constant & stiffness scale \\
$t_c$ & $t_c=\sqrt{m_c/k_c}$ & time scale \\
$\hat{m}$ & $m_b/m_c$ & ball-to-chain mass ratio \\
$\hat{k}$ & $k_b/k_c$ & ball-to-chain spring ratio \\
$\hat{v}$ & $v_0/\bigl(d_c\sqrt{k_c/m_c}\bigr)$ & dimensionless impact velocity \\
$\hat{g}$ & $g\,m_c/(k_c d_c)$ & dimensionless gravity \\
$\hat{\zeta}$ & $\zeta/\sqrt{k_c m_c}$ & dimensionless chain damping \\
$N$ & $N$ & number of particles \\
\bottomrule
\end{tabular}
}%
\end{table}

The equations of motion are integrated with a velocity-Verlet scheme, with a time step fixed as a fraction of the ball's natural period.

\subsection{Parameter choices and measurement procedures}

Parameters are chosen near the laboratory regime while keeping the model simple. We set $d_c=m_c=k_c=1$, so $m_b=\hat{m}$ and $k_b=\hat{k}$; chain and ball damping are equal, $\zeta=\zeta$. The ball-to-chain mass ratio is $\hat{m}=20$, so the impactor is much heavier than the cells it strikes. The dimensionless impact velocity is $\hat{v}=0.1$ ($10\%$ of the characteristic spring energy $k_c d_c^2$), matching laboratory conditions and avoiding nonphysical tunneling through particles. Gravity is set to $\hat{g}=0$: it does not measurably affect restitution or contact time, and removing it eliminates poorly constrained pre-loading, rebound-arch corrections, and variable drop-height parameters, isolating the timing competition between wave return and contact. With $\hat{g}=0$ the springs carry no pre-load, as in the experiment, and the ball's contact time is set by its own inertia rather than by gravity.

Without gravity holding the ball against the stack, the contact can open briefly even while the ball is still driving compression, so rebound must be defined carefully. We take $t_{\rm impact}$ as the first time step at which the ball--chain center-to-center distance drops below one diameter $d_c$, and $t_{\rm rebound}$ as the first time step after $t_{\rm impact}$ at which that distance exceeds $1.9\,d_c$ -- one diameter plus a $0.9\,d_c$ buffer, so that brief contact breaks during compression are not mistaken for rebound. Because $\hat{g}=0$, the ball does not accelerate in free flight, so the exact buffer size affects only measurement precision, not the result. The simulated restitution is $e_{\rm sim}=|v_{\rm rebound}|/|v_{\rm impact}|$, with $v_{\rm impact}$ the ball velocity at first contact and $v_{\rm rebound}$ its velocity immediately after $t_{\rm rebound}$.

The wave speed $c_{\rm sim}$ is measured as in the experiment: from the one-way travel time between first contact and the first significant compression of the last particle,
\begin{equation}
c_{\rm sim} = \frac{(N-1)d_c}{t_{\rm travel}}.
\end{equation}
We then define
\begin{equation}
\Lambda_{\rm sim}(N) = \frac{2(N-1)d_c}{c_{\rm sim}\,\tau_{\rm theory}},
\end{equation}
with the contact time approximated by a quarter-period of the impactor's spring--mass oscillation,
\begin{equation}
\tau_{\rm theory} = \frac{\pi}{2}\sqrt{\frac{m_b}{k_b}}.
\end{equation}
Direct contact-time measurements from the simulation are unreliable, since $\hat{g}=0$ allows brief, non-physical contact breaks during compression; $\tau_{\rm theory}$ avoids this but likely underestimates the true contact duration, especially for stiffer ball--chain contacts. As a result $\Lambda_{\rm sim}$ does not peak exactly at $\Lambda=1$, even though the underlying mechanism is the same timing match between wave return and contact end. A more accurate numerical estimate of $\tau$ is left for future work.

Normalized restitution is defined as in the experiment: for each parameter set $(\hat{m},\hat{k},\hat{\zeta},\hat{v},\hat{g})$ we take $e_{\max}$ as the peak of $e_{\rm sim}(N)$ and $e_\infty$ as the average of the last three points at large $N$, then plot
\begin{equation}
\tilde{e}_{\rm sim} = \frac{e_{\rm sim}-e_{\infty}}{e_{\max}-e_{\infty}}
\end{equation}
against $\Lambda_{\rm sim}$ for comparison with the experimental master curve (main text, Fig.~4).

\subsection{Results}

\begin{figure}[t]
\includegraphics[width=.5\columnwidth]{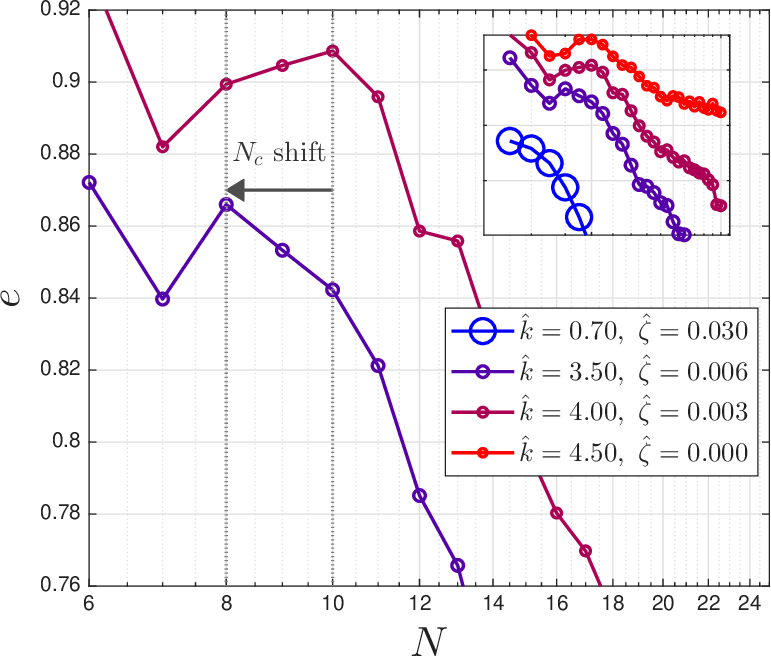}
\caption{
    Coefficient of restitution from the idealized one-dimensional ball--chain simulation plotted against chain length $N$ for several values of the ball--chain spring ratio $\hat{k}$ and damping $\hat{\zeta}$.
    Decreasing the effective stiffness (and increasing the corresponding damping) shifts the resonance maximum to lower $N$, reproducing the non-monotonic $e(N)$ behavior and pressure-dependent shift of $N_c$ seen in the paper-stack experiment.
    All simulations use $\hat{m}=20$, $\hat{v}=0.1$, and $\hat{g}=0$, placing the model in the same regime of mass ratio, impact velocity, and negligible gravity as the laboratory experiments.
    }
    \label{fig:simCorrVsN}%
\end{figure}

For Fig.~\ref{fig:simCorrVsN} we sweep $N=3$ to $30$ at fixed $\hat{m}=20$, $\hat{v}=0.1$, $\hat{g}=0$, over four spring--damping pairs $(\hat{k},\hat{\zeta})$ spanning soft/strongly-damped to stiff/weakly-damped stacks. The main panel shows the two intermediate pairs, where the resonance peak is clearest; the inset shows all four.

As in the experiment, $e_{\rm sim}$ falls at small $N$, recovers to a maximum at intermediate $N$, then decays once the chain behaves as a semi-infinite absorber. The maximum shifts with $\hat{k}$ and $\hat{\zeta}$ in the same sense as the experimental shift of $N_c$ with paper grade and ambient pressure (main text, Figs.~1 and 3): lowering $\hat{k}$ and raising $\hat{\zeta}$ softens the cells and increases their dissipation, lowering the wave speed and moving the resonance to smaller $N$ -- the same direction as lowering pressure or changing paper grade.

That this curve emerges from an idealized 1D model shows that the non-monotonic $e(N)$ is a resonance effect, not a detail of the paper's microstructure. The model also isolates which parameters set the shape of $e(N)$ under pressure, and points to the coupled stiffness--damping mechanism as the next step toward connecting $e(N)$ to the medium's microstructure.

\subsection{Mass-ratio resonance and the bar-impact baseline}

\begin{figure}[t]
\includegraphics[width=.5\columnwidth]{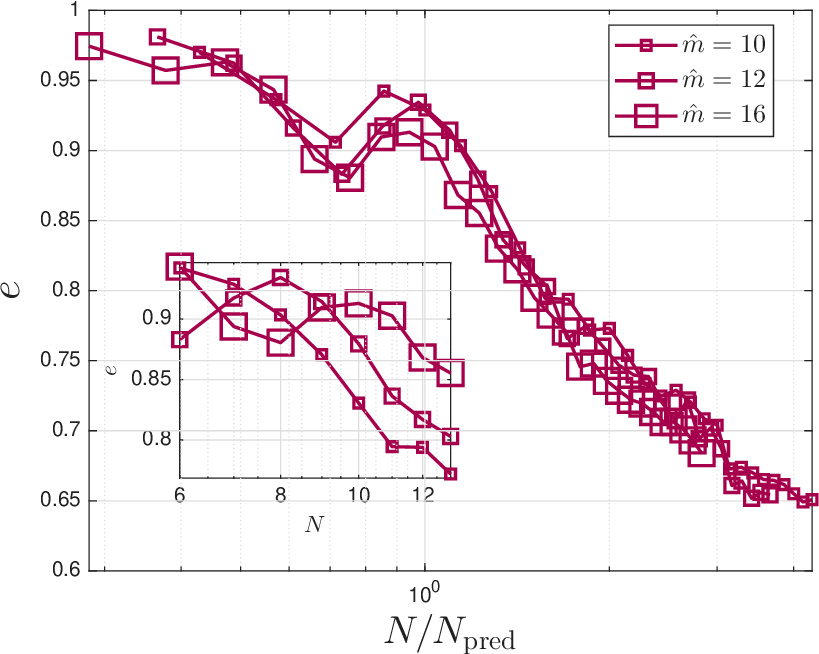}
\caption{
    Mass-ratio resonance in the idealized one-dimensional ball--chain simulation.
    Main panel: coefficient of restitution $e_{\rm sim}$ plotted versus the normalized chain length $N/N_{\rm pred}$ for three ball--to--chain mass ratios $\hat{m}=10$, 20, and 40, at fixed $\hat{v}=0.1$, $\hat{g}=0$, and $(\hat{k},\hat{\zeta})$.
    As $\hat{m}$ increases, the resonance peak shifts to larger $N$, and the peak positions satisfy $(N_{\rm peak}-1)\simeq0.62\,\hat{m}$, as predicted by the Saint-Venant bar-impact condition $M_s/m_b\simeq\pi^2/16$.
    The inset shows the same data without normalization, $e_{\rm sim}(N)$, illustrating directly how well the normalized curves collapse and highlighting that the recovery in $e_{\rm sim}$ is only a few percent, much smaller than in the experiment, so timing in an elastic bar accounts for where the maximum lies but not for its size.
}
    \label{fig:massRatioEffectSim}%
\end{figure}

The ball--chain model is a discretized longitudinal bar where the total chain mass $M_c=(N-1)\, m_c$ plays the role of the bar, and the ball mass $m_b=\hat{m} \, m_c$ acts as the impactor.
The Saint-Venant bar-impact solution predicts that the
timing parameter

$$
\Lambda = \frac{t_r}{\tau} \simeq \frac{4}{\pi}\, \sqrt{\frac{M_c}{m_b}}
$$
attains $\Lambda\simeq1$ when $M_s/m_b\simeq\pi^2/16\approx0.62$.
In the chain this translates to
$$
N-1 \simeq 0.62\,\hat{m},
$$
so the resonance peak in $e_{\rm sim}(N)$ should move linearly with $\hat{m}$.

Figure~\ref{fig:massRatioEffectSim} tests this directly by sweeping $\hat{m}$ at fixed $(\hat{k},\hat{\zeta},\hat{v},\hat{g})$.
The peak positions follow $(N_{\rm peak}-1)\simeq0.62\,\hat{m}$, confirming that the 1D model realizes the classical bar-impact mass-ratio resonance.
The associated increase in $e_{\rm sim}$ is only a few percent, however, far smaller than the $\sim80\%$ recovery seen in the paper-stack experiment.
The ball--chain simulation therefore serves as a bar-theory baseline and explains why the maximum in $e(N)$ occurs near the observed stack-to-ball mass ratio, while highlighting that additional channels (trapped air, coupled stiffness and damping) are required to account for the size and pressure sensitivity of the effect. 

\subsection{Effective stiffness, damping, and pressure: guidance for future work}
\label{sec:PressureMapping}

How pressure sets the effective stiffness and damping is not known; a microscopic model of this is beyond the present scope. The 1D chain gives an initial probe: lowering $\hat{k}$ while raising $\hat{\zeta}$ reproduces low-pressure curves, and raising $\hat{k}$ while lowering $\hat{\zeta}$ reproduces high-pressure ones, matching the experimental trends under pressure and paper-grade changes (main text, Figs.~2--4), even though the mapping from $(\hat{k},\hat{\zeta})$ to $P_0$ is not known.

This suggests pressure tunes stiffness and damping together rather than either alone, and that this coupling is enough to reproduce the observed shifts in the resonance peak and plateau. The functional form of that dependence, and its connection to the paper--air microstructure, is left for future work.

\end{document}